\documentclass[multphys,vecphys]{svmult}

\usepackage{makeidx}     
\usepackage{graphicx}    
\usepackage{multicol}    

\makeindex             

\begin{document}

\title*{Disappearance of N$_2$H$^+$ from the Gas Phase in the Class 0 Protostar IRAM~04191}
\titlerunning{Disappearance of N$_2$H$^+$ from the Gas Phase in IRAM~04191}
\author{A. Belloche\inst{1,2}\and
P. Andr\'e\inst{3}}
\institute{Max-Planck-Institut f\"ur Radioastronomie, Auf dem H\"ugel 69, 
           D-53121 Bonn, Germany,
           \texttt{belloche@mpifr-bonn.mpg.de}
\and LRA/LERMA, Ecole Normale Sup\'erieure, 24 rue Lhomond, 
     F-75231 Paris Cedex 05, France
\and Service d'Astrophysique, CEA/Saclay, F-91191 Gif-sur-Yvette, France,
     \texttt{pandre@cea.fr}}
%
%
\maketitle

 {\bf Abstract}. We present a high-resolution spectroscopic study of the
 envelope of the young Class~0 protostar IRAM~04191+1522 in 
 Taurus. N$_2$H$^+$(1-0) observations with the Plateau de Bure Interferometer 
 and the IRAM 30m telescope demonstrate that the molecular ion N$_2$H$^+$ 
 disappears from the gas phase in the inner envelope ($r <$ 
 1600 AU, $n_{\mbox{\tiny {H$_2$}}} > 4-7 \times 10^5$ cm$^{-3}$).
 This may result from N$_2$ depletion on polar ice mantles and enhanced 
 grain chemistry.

\section{The Class 0 Protostar IRAM~04191}
\label{sec_iram04191}

With an age $t \sim 1 -3 \times 10^4$~yr measured from the beginning of the 
accretion phase, IRAM~04191+1522 is the youngest Class 0 protostar known so 
far in the Taurus molecular cloud ($d = 140$ pc). It has a prominent envelope 
(1.5 M$_\odot$) and an associated powerful bipolar outflow \cite{Andre99}.
Belloche et al. \cite{Belloche02} showed that the envelope is undergoing 
both extended infall motions and fast, differential rotation. They proposed 
that the rapidly rotating inner part of the envelope ($r < 3500$ AU) 
corresponds to a magnetically supercritical core decoupling from an 
environment still supported by magnetic fields and strongly affected by 
magnetic braking.
   
We recently carried out new observations in the N$_2$H$^+$(1-0) line with the 
Plateau 
de Bure Interferometer (PdBI) to probe the inner structure of the IRAM~04191 
envelope. Here, we present and discuss the results of these
high-resolution ($\sim 5''$) observations of the small-scale distribution 
of N$_2$H$^+$.

\section{Small-Scale N$_2$H$^+$(1-0) Emission}
\label{sec_n2h+_pdbi}

\begin{figure}[!ht]
\centering
  \resizebox{\hsize}{!}{\includegraphics[angle=270]{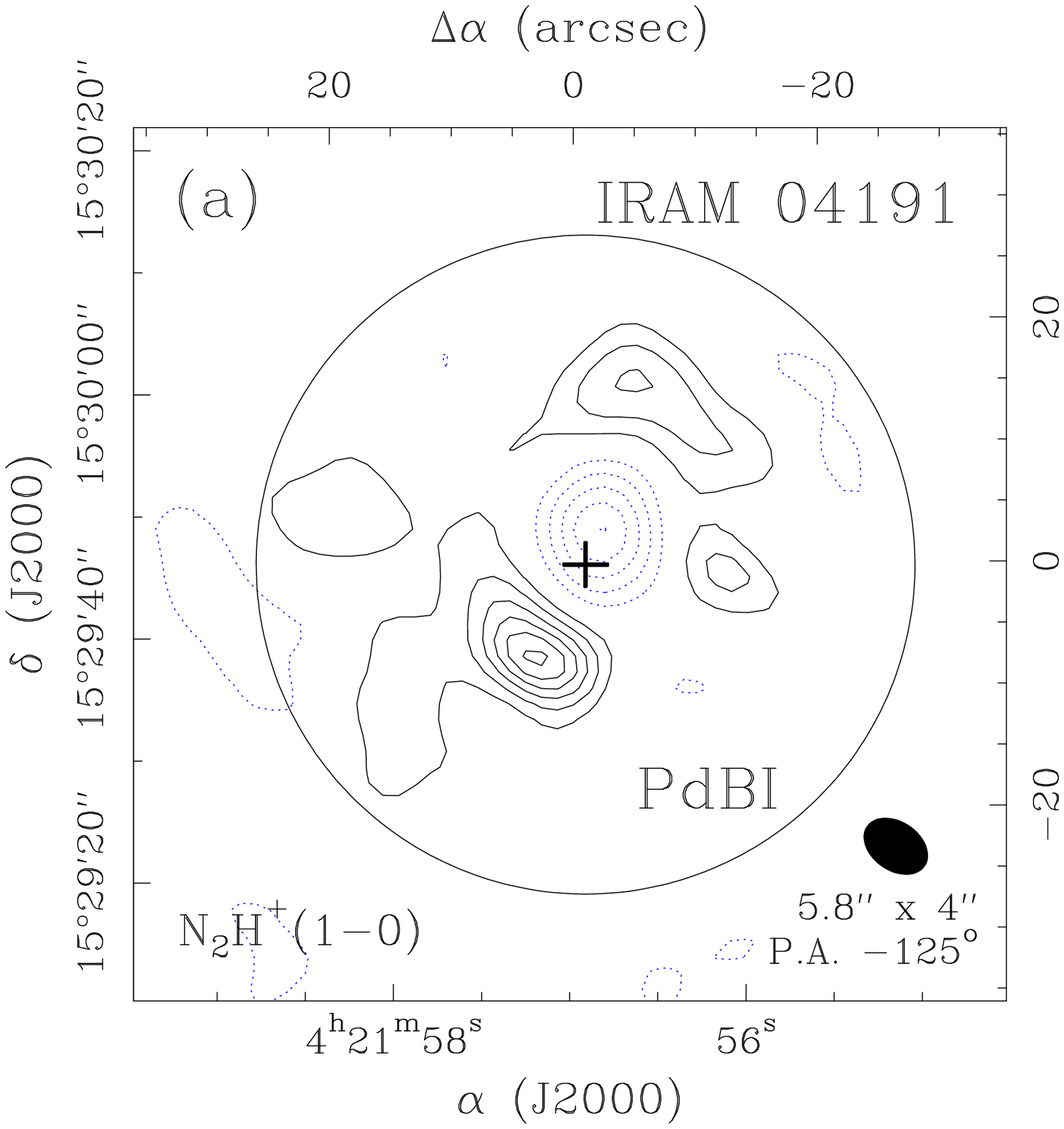}\hspace*{0.5cm}\includegraphics[angle=270]{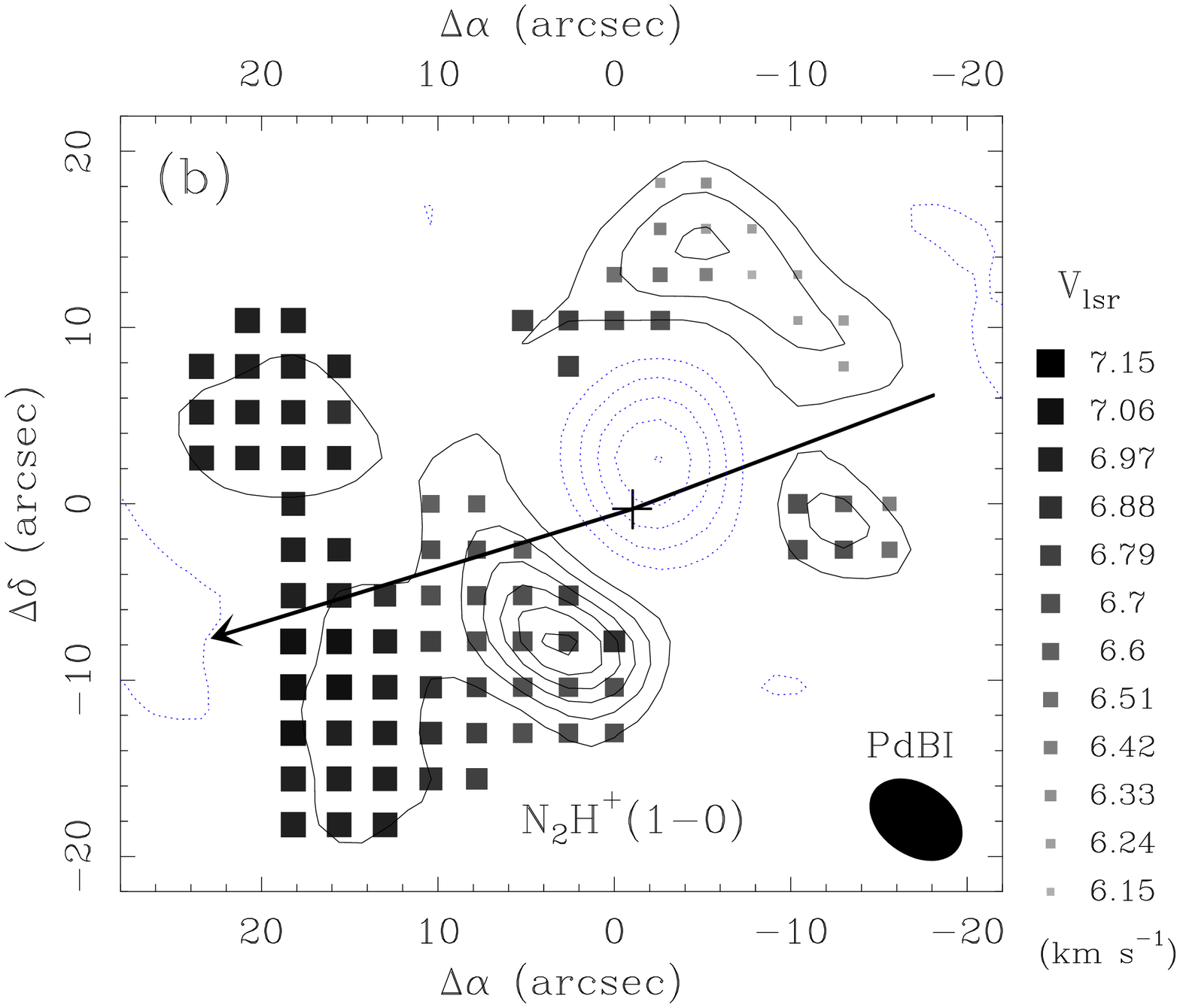}}
\caption{\textbf{(a)} N$_2$H$^+$(1-0) integrated intensity map of the 
 IRAM~04191 envelope obtained with PdBI. The contour step is 0.1 
 Jy~beam$^{-1}$~km~s$^{-1}$ (i.e. 3 times the rms noise). The big circle
 represents the primary beam. The emission gap very close to the central 
 position ({\it black cross}) suggests a strong decrease of the N$_2$H$^+$ 
 abundance in the inner part of the envelope. 
 \textbf{(b)} Map of the LSR velocity 
 ({\it squares}) as derived from Gaussian hfs fits to the N$_2$H$^+$(1-0)
 spectra. The map of integrated intensity is overlaid as contours.
 We measure a mean velocity gradient of $\sim$ 26 km~s$^{-1}$~pc$^{-1}$ across
 the envelope ({\it black arrow}), whose direction at PA $\sim 107^\circ$ is 
 consistent with the rotation seen on larger scale.
 }
\label{fig_n2h+_pdbi}
\end{figure}

Fig.~\ref{fig_n2h+_pdbi}a shows the N$_2$H$^+$(1-0) integrated intensity map 
obtained with PdBI, which filters out the extended emission observed with the 
IRAM 30m telescope. The (naturally-weighted) synthesized half-power beamwidth 
is $5.8''\times 4''$. Two emission peaks are seen approximately along the
direction perpendicular to the outflow axis, on either side of a
gap very close to the central position. This suggests a strong decrease of 
the N$_2$H$^+$ abundance toward the center. Besides, the two emission peaks 
have very different centroid velocities (see Fig.~\ref{fig_n2h+_pdbi}b): we 
measure a strong velocity gradient of $\sim 26$ km~s$^{-1}$~pc$^{-1}$ along 
the axis with PA = 107$^\circ$, i.e. approximately the same axis 
along which a rotational velocity gradient is detected 
on larger scale \cite{Belloche02}. The gradient measured with PdBI is 
8 times larger than the mean velocity gradient observed with the 30m 
telescope toward the core, which confirms the presence of fast, 
differential rotation in the IRAM~04191 envelope.

\section{N$_2$H$^+$ Disappearance from the Gas Phase}
\label{sec_n2h+_pdbi30m}

\begin{figure}[!ht]
\centering
  \resizebox{\hsize}{!}{\includegraphics[width=0.50\linewidth,angle=270]{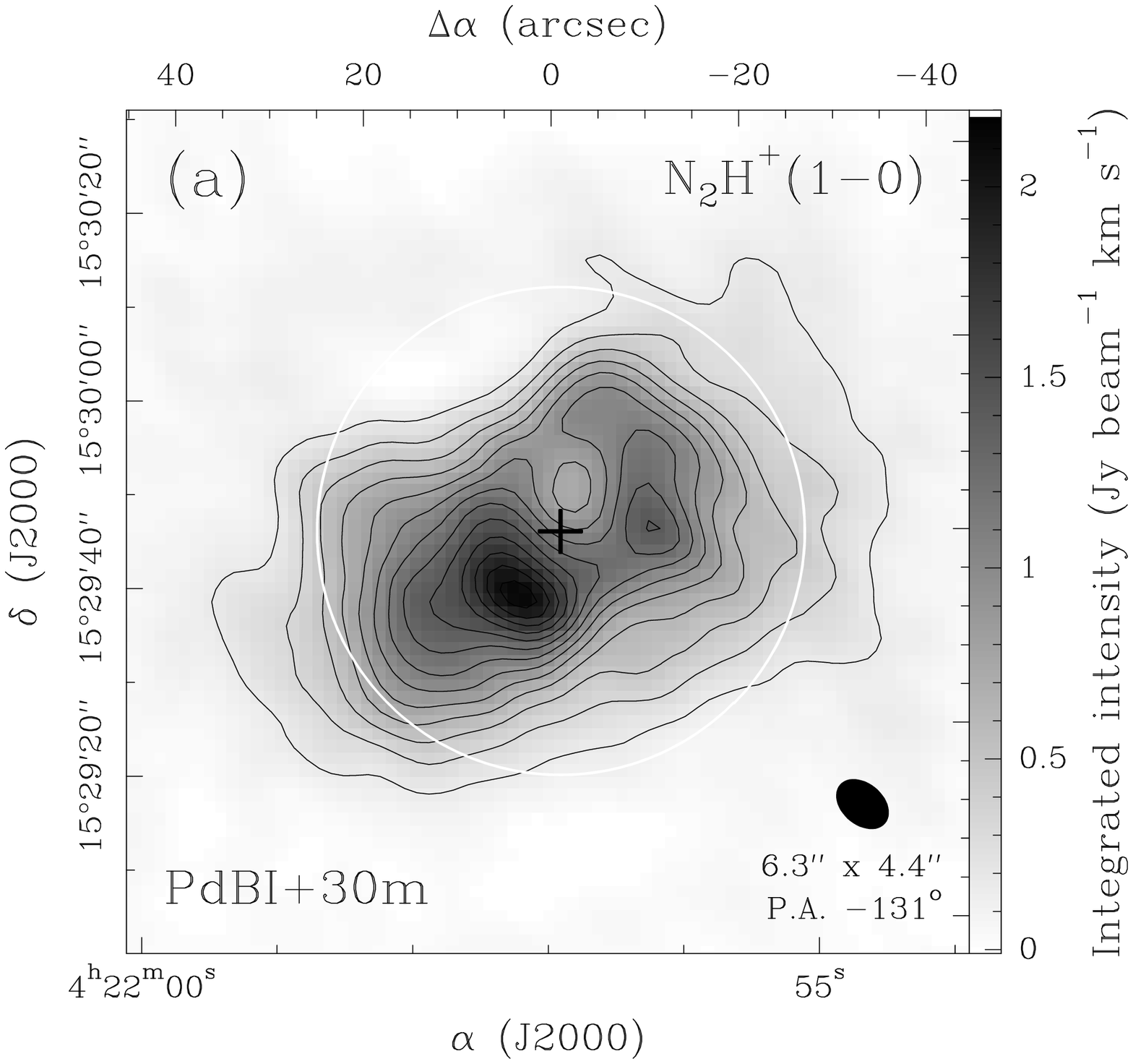}\hspace*{0.5cm}\includegraphics[width=0.50\linewidth,angle=270]{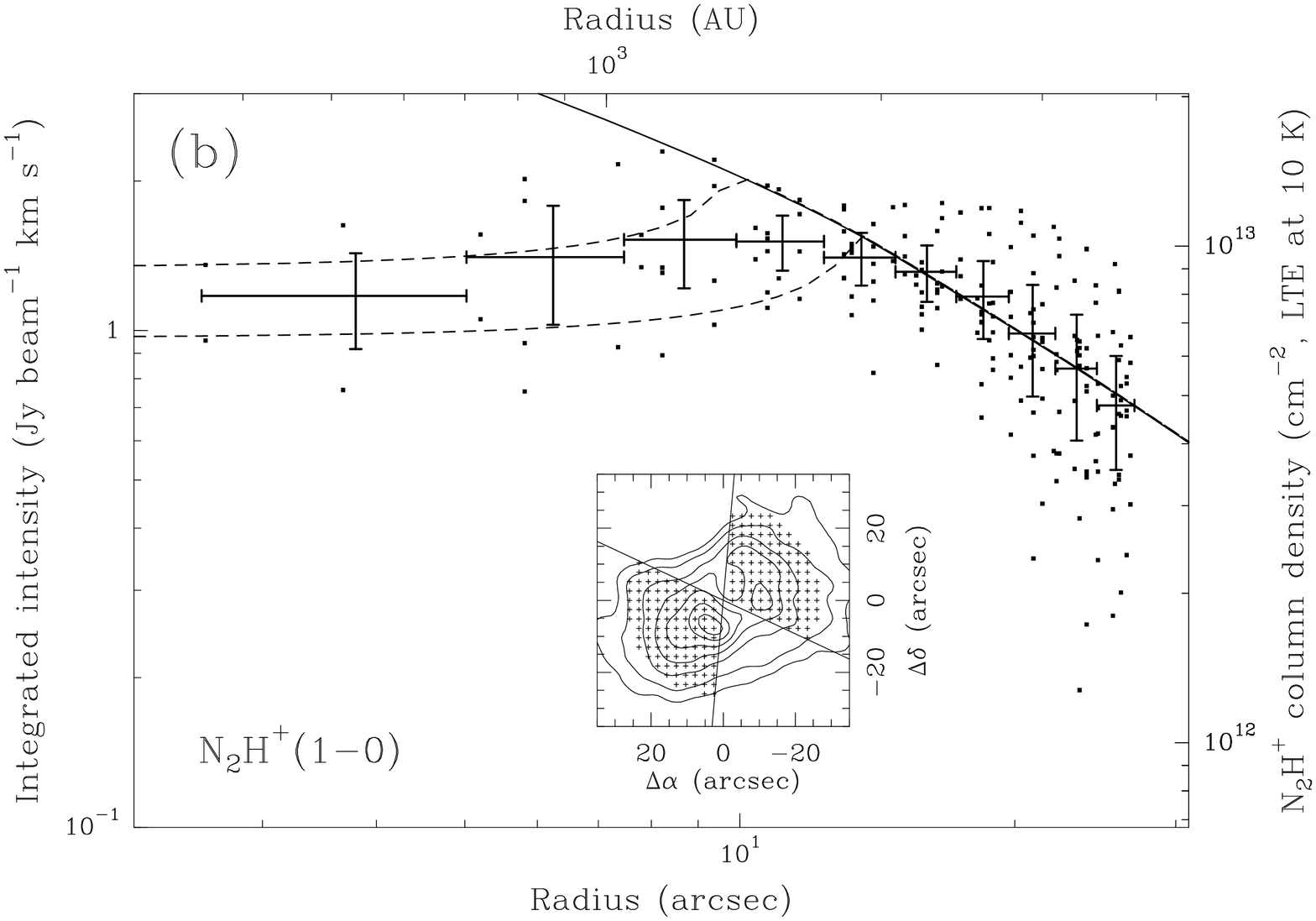}}
\caption{\textbf{(a)} N$_2$H$^+$(1-0) integrated intensity map of the 
 IRAM~04191 envelope combining the PdBI data and the short spacings obtained 
 at the 30m telescope. The contour step is 0.16 Jy~beam$^{-1}$~km~s$^{-1}$ 
 (i.e. 7 times the rms noise). The white circle represents the PdBI primary 
 beam.
 \textbf{(b)} Radial profile of the 
 N$_2$H$^+$(1-0) integrated intensity corrected for the primary beam 
 attenuation. The dots with error bars represent the average intensity in 
 concentric shells limited to the positions shown in the insert. The solid 
 curve indicates the H$_2$ column density profile \cite{Motte01,Belloche02} 
 scaled to the 9$^\mathrm{th}$ shell for comparison. The dashed curves show 
 the effect of a hole of N$_2$H$^+$(1-0) emission with radius r = 1350 and 
 1850 AU, respectively. They suggest a strong decrease of the N$_2$H$^+$ 
 abundance in the inner part of the envelope ($r < 1600$ AU).}
\label{fig_n2h+_pdbi30m}
\end{figure}

Since the interferometer filters out the extended emission, we added short 
spacings obtained with the 30m telescope \cite{Belloche02} to the PdBI data. 
As shown in the N$_2$H$^+$(1-0) integrated intensity map (see 
Fig.~\ref{fig_n2h+_pdbi30m}a), the emission is resolved in a ringlike structure
with two local peaks, surrounding a gap close to the center. Gaussian hfs fits 
to the 7-component N$_2$H$^+$(1-0) spectra yield a mean total opacity 
$\tau_\mathrm{tot} = 10 \pm 4$ in the central part ($\theta < 10''$), 
corresponding to $\tau = 2.6 \pm 1.0$ for the strongest component and 
$\tau = 1.1 \pm 0.4$ for the isolated component. However, the opacity map does 
not show any enhancement toward the center and the gap close to the center 
does not have the highest opacity. Besides, the intensity maps integrated over
each $\Delta F_1$ group of N$_2$H$^+$(1--0) hyperfine
components separately show the same pattern. We are 
thus confident that the observed emission gap does not result from an opacity 
effect.

Assuming a constant excitation temperature for  N$_2$H$^+$(1-0), we
compare in Fig.~\ref{fig_n2h+_pdbi30m}b the circularly-averaged profile of 
the N$_2$H$^+$(1-0) integrated intensity, corrected for PdBI primary beam 
attenuation, with the H$_2$ column density profile derived from the dust 
emission \cite{Motte01,Belloche02}. The N$_2$H$^+$(1-0) profile departs from 
the H$_2$ column density profile for $\theta \leq 13''$. A good model fit is 
obtained when one assumes a hole of N$_2$H$^+$(1-0) emission in the central 
part of the envelope ($r <$ 1350-1850 AU, see dashed curves). This crude
comparison thus strongly suggests that the N$_2$H$^+$ ion disappears from the
gas phase  above a density of $n_{\mbox{\scriptsize H$_2$}} \sim 4-7 \times 
10^5$ cm$^{-3}$ in the IRAM~04191 envelope.

\section{Discussion}
\label{sec_discuss}

Very little evidence for N$_2$H$^+$ abundance drop has been reported so far in 
dense cores (see however the case of B68 \cite{Bergin02}). The N$_2$H$^+$(1-0) 
emission usually follows the dust emission very well 
in prestellar cores 
and has long been thought to be one of the best molecular tracers of dense gas 
\cite{Tafalla02,Caselli02}. From a theoretical point of view, all recent 
chemical models predict a centrally-peaked N$_2$H$^+$ column density profile 
during the prestellar collapse phase 
\cite{Bergin97,Aikawa01,Aikawa03,Li02,Shematovich03}. In particular, the 
models calculated  by Aikawa et al. \cite{Aikawa03} including gas-phase 
reactions, gas-dust interactions and diffusive grain-surface reactions predict 
an increase of the N$_2$H$^+$ abundance during the core collapse up to 
$n_{\mbox{\scriptsize H$_2$}} \sim 1 \times 10^7$ cm$^{-3}$ or more. Since 
only a few $10^4$ yr have elapsed and the density at $r \sim 1600$ AU cannot 
have changed much between point mass formation (the last stage of this model)
and the present stage of IRAM~04191, 
the drop observed in the N$_2$H$^+$ abundance above 
$n_{\mbox{\scriptsize H$_2$}} \sim 4-7 \times 10^5$ cm$^{-3}$
is strongly at variance with current theoretical predictions. 

Our result implies either a stronger dependence of the N$_2$H$^+$ abundance on 
density (i.e. an earlier drop in abundance during the prestellar 
phase) or an abundance drop on a timescale as short as a few $10^4$ yr.
Since N$_2$ is the precursor of N$_2$H$^+$, the drop in N$_2$H$^+$ abundance
may result from stronger N$_2$ depletion onto grain surfaces. This could 
happen if N$_2$ were depleted on H$_2$O ice mantles characterized by a higher
binding energy (see \cite{Bergin02}, Model C of \cite{Aikawa01} and 
Fig.~6 of \cite{Shematovich03}) and/or if the chemistry transforming  
N$_2$ in less volatile species on grain surfaces were enhanced. 
It is also possible that N$_2$H$^+$ destroyers such as CO have returned to
the gas phase in the inner envelope, following heating by the central 
protostar or shock by the outflow. However, the dust temperature is below 10 K 
at $r \sim 1000$ AU and our C$^{18}$O(2-1) observations \cite{Belloche02} at a 
resolution of 800 AU in radius show depletion toward the envelope center by a 
factor of 3 in column density when compared to the 1.3mm continuum emission. 
This latter explanation seems thus less likely.

A better tracer than N$_2$H$^+$ will thus have to be found to probe the
velocity structure of cold protostellar envelopes at high resolution with the
next generation of (sub)millimeter instruments such as ALMA.
H$_2$D$^+$ may be such a good tracer \cite{Bergin02,Caselli03}.

\vspace*{1.ex}
\noindent {\footnotesize \textit{Acknowledgements.} We would like to thank 
Y. Aikawa, E. Bergin, H. Fraser, M. Gerin, D. Johnstone, G. Pineau des 
For\^ets and M. Walmsley for enlightening discussions on chemistry, and 
F. Gueth, R. Neri and J. Pety for their help with the reduction of our PdBI 
observations.}
\vspace*{-1ex}

\printindex
\end{document}